\documentclass[aps,prfluids,onecolumn,10pt]{revtex4-2}
\usepackage{amsmath}
\usepackage{graphicx}
\usepackage{hyperref}
\usepackage{amssymb}
\usepackage{wrapfig}
\usepackage[dvipsnames]{xcolor}
\usepackage[normalem]{ulem}

\usepackage{geometry}
\begin{document}

\title{Transport of spherical microparticles in a 3D vortex flow}

\author{Marine Aulnette$^{1\dagger}$}
\author{Noa Burshtein$^{1\dagger}$}
\author{Arash Alizad Banaei$^2$}
\author{Luca Brandt$^3$} 
\author{Simon J. Haward$^4$}
\author{Amy Q. Shen$^4$}
\author{Blaise Delmotte$^5$}
\author{Anke Lindner$^{1,6*}$}
\affiliation{$^1$Laboratoire de Physique et Mécanique des Milieux Hétérogènes (PMMH), ESPCI Paris, PSL University, CNRS, Sorbonne University, and Paris Cité University, 75005 Paris, France}
\affiliation{$^2$PDC, KTH}
\affiliation{$^3$Department of Environment, Land and Infrastructure Engineering, Politecnico di Torino, Torino, Italy}
\affiliation{$^4$Okinawa Institute of Science and Technology, Graduate University, Okinawa, Japan}
\affiliation{$^5$LadHyX, CNRS, Ecole Polytechnique, Institut Polytechnique de Paris, 91120 Palaiseau, France}
\affiliation{$^6$Institut Universitaire de France, Paris, France}

\thanks{$^{\*}$Corresponding author}
\thanks{$^\dagger$ The authors contributed equally. }

\date{\today}

\begin{abstract}
\vspace{0.4 cm}
Particles are common in biological and environmental flows and are widely used in industrial and pharmaceutical applications. Their motion and flow dynamics are strongly affected by interactions with the surrounding flow
structure. While particle-flow interactions have been extensively studied in low Reynolds number (Re) flows as well as in fully developed turbulence, the transport mechanisms of these particles in intermediate flow regimes remain
less explored. Here, we investigate the response of neutrally buoyant spherical particles to a single vortex flow field. Using a microfluidic cross-slot geometry, we generate a well-characterized, stationary, three-dimensional streamwise vortex at moderate $\text{Re}$ ($\sim 50$). Our experimental results, supported by numerical simulations, show that with increasing particle diameter, they are progressively excluded from the vortex core. Initially, small particles follow a Burgers vortex-like self-similar motion, but for larger particle diameters, deviations from this trend emerge due to fluid inertia and finite-size effects. These findings enhance our understanding of particle dynamics in vortical flows and have implications for microfluidic applications involving particle sorting and separation.
\end{abstract}

\maketitle

\section{Introduction}
Exploring particle-vortex interactions, where small particles interact with swirling vortical structures, provides valuable insights for optimizing microfluidic devices~\cite{burshtein2019, park2019, tanasijevic2022}, advancing environmental monitoring techniques~\cite{wang2020, zhang2022}, and refining particle separation methods~\cite{volpe2019, nasiri2020Review}. A comprehensive understanding of these interactions also contributes to the broader field of fluid mechanics phenomena, including turbulence and mixing, cloud formation, precipitation, and climate dynamics~\cite{shaw2003}.
In particular, particle-vortex interactions play a pivotal role in the design of sorting and separation techniques, which are increasingly prominent in both applied and fundamental research. These devices exploit geometric constraints and non-linear hydrodynamic forces to facilitate passive sorting based on the physical properties of the particle~\cite{tang2022}.

In microfluidic applications, such non-linear hydrodynamic forces may arise from fluid inertia~\cite{dicarlo2009Inertial_microfluidics}, visco-elasticity~\cite{seo2014, d2017particle}, or particle deformability~\cite {secomb2017blood}. In this study, we focus on the migration of rigid spherical particles induced by inertial effects. Inertial microfluidics enables precise particle manipulation in an intermediate flow regime, where inertial forces shape the fluid structure and particle dynamics~\cite{dicarlo2009Inertial_microfluidics}. 
During the 1960s, Segré and Silberberg demonstrated the radial migration of rigid, neutrally buoyant spherical particles toward specific positions in a tube for Reynolds numbers ($\text{Re}$, the ratio between inertial to viscous forces) in the range  $10 \lesssim $ \text{Re} $ \lesssim 100$. These conditions correspond to particle Reynolds numbers $\text{Re}_p$ of order one or below~\cite{segre1961radial,segre1962behaviour}. 
Other studies have considered migration in the presence of external forces, like gravity, within a similar range of Reynolds numbers. For such non-neutrally buoyant particles, a slip velocity between the fluid and the particle along the particle streamlines can be externally induced. This slip generates so-called Saffman lift forces, which act perpendicular to the flow direction~\cite{stone2000philip} and drive particle migration across streamlines. A similar cross-streamline migration can arise from the difference between the particle's rotation rate and local fluid vorticity and can analogously lead to cross-streamline migration due to the Magnus effect~\cite{matas2004inertial,rubinow1961transverse}. 
Most studies of the nature of inertial lift force have focused on straight, rectangular channels. In these configurations, a combination of shear-induced lift and wall-induced repulsion drives particles towards specific equilibrium positions~\cite{stone2000philip,dicarlo2009PRL}. Although typical migration velocities are small compared to the mean flow velocity,  their cumulative effect along particle trajectories is visible before equilibrium positions are reached. Mathematical modeling of inertial lift forces relies on perturbation methods that are valid in the regime of intermediate $\text{Re}$ (typically corresponding to small Re$_p$). However, such models are mostly restricted to simple unidirectional flows ~\cite{,schonberg1989inertial,asmolov1999,hood2015inertial}, with the exception of a recent study on particle migration in Taylor-Couette flows \cite{davoodi2024sedimenting}, limiting their applicability in the design of more complex inertial separation devices.

A few investigations have considered secondary flows arising in addition to the primary unidirectional flows due to the presence of channel curvature. Such secondary flows originate from centrifugal forces and typically occur in the form of two vertically stacked, counter-rotating vortices of small magnitude, or more complex vortical structures~\cite{dean1928fluid}. This secondary flow exerts an additional force on the particles, known as Dean drag. The interplay between inertial forces and Dean drag determines distinct particle focusing positions~\cite{di2007continuous}.

Here, we investigate experimentally and numerically the inertial migration of rigid spherical particles within a single three-dimensional vortex. Such a vortex can serve as a fundamental element in microfluidic separation devices, and also arise in large-scale flows, such as 3D turbulence~\cite{brandt2022particle}. The vortex flow features spatially varying shear and vorticity, and particles are transported along curved streamlines. Our focus is on intermediate $\text{Re}$ regimes, which remain largely unexplored and for which no theoretical framework is currently available. 
In addition, we consider the impact of finite-size effects, which emerge when particle dimensions become comparable to spatial variations in the flow field.

To study particle interactions in a simplified single vortex, a steady-state streamwise vortex can be generated using a cross-slot flow configuration (see Figure~\ref{fig:figure1}). Fluid flow in the cross-slots junction is complex and has been extensively investigated. A three-dimensional spiral vortex develops in the cross-slot resulting from a symmetry-breaking flow bifurcation above a critical $\text{Re}$ determined by the aspect ratio. The onset of this instability along with the structure and temporal fluctuations of the vortex, has been characterized as a function of the aspect ratio of the device and the elasticity of the fluid~\cite{haward2016tricritical,burshtein2017inertioelastic,burshtein2019controlled,burshtein2021periodic}. Such a configuration is ideal for our study as it is well characterized and allows clear experimental observations in several planes of interest.

Although several studies have investigated fluid dynamics and particle trapping in cross-slots junctions~\cite{zhang2022capturing,gossett2012hydrodynamic,tse2013quantitative,tanyeri2011microfluidic}, few have considered moderate flow inertia, neutrally buoyant particles, and precise shapes of trajectories, particularly in stretched vortices. One such study numerically simulated large, rigid spherical particle behavior in a cross-slot geometry at a relatively high Re (Re = 120)~\cite{kechagidis2023numerical}. These simulations explored how particle trajectories vary with different initial positions and confinement levels, focusing on large particles and exploring a high confinement regime.

In this paper, we present an experimental and numerical study of finite-size neutrally buoyant spherical particles transported in a three-dimensional isolated, steady, and well-controlled vortex within a microfluidic system. By focusing on small particle sizes with relatively low confinement (between $2\%$ and $10\%$), and exploring moderate inertia regimes, the study addresses an important gap in the literature. Flow characterization and full three-dimensional particle dynamics are investigated experimentally in two orthogonal planes of observation. Numerical simulations using immersed boundary methods complement these experiments by providing a complete representation of the base flow field, fluid flow parameters, and particle trajectories. 
This study investigates how finite-size effects and inertia influence the trajectories of neutrally buoyant spherical particles in a controlled vortical flow. A comprehensive understanding of these mechanisms is crucial for refining predictive models of particle dynamics in microfluidic and vortical flow systems. Additionally, our study will provide reference data for point-particle or discrete-element methods that can follow the particle dynamics at significantly lower cost than the interface-resolved simulations used here.
Beyond fundamental fluid dynamics, this study has broad implications for applications in biology and engineering, particularly in particle sorting, trapping, and active particle multiplication~\cite{vigolo2014unexpected, hur2011high, wang2015size,thouvenot2024high,kang2020intracellular}. The ability to characterize stress history and predict particle trajectories will improve efficiency and outcome for these applications.

The paper is structured as follows. The experimental set-ups and the numerical method are presented in section~\ref{sec:setup}. Careful characterization of the base flow is performed with comparison between $\mu$-PIV measurements and simulated velocity and vorticity fields in section~\ref{sec:PIV}. Then, particle dynamics is discussed in section~\ref{sec:particle_dyn}. Our joint experimental and numerical approaches are complementary, as they respectively allow access to distinct statistical and specific dynamical properties of both the flow and the particle trajectories. As a first step, statistics of the number of trajectories in the vertical mid-plane are explored. Then, we focus on individual particle trajectories, using both experimental and numerical data. Finally, we discuss finite size and inertia effects on the particles using specific properties of the base flow along particle trajectories, as well as microscopic particle observables such as velocity and rotation. The paper concludes with a summary of the main findings in Section~\ref{sec:conclusion}.

\section{Methods} \label{sec:setup}

\subsection{Experimental set-up}

\begin{figure}[t]
\centering
\includegraphics[width=7cm]{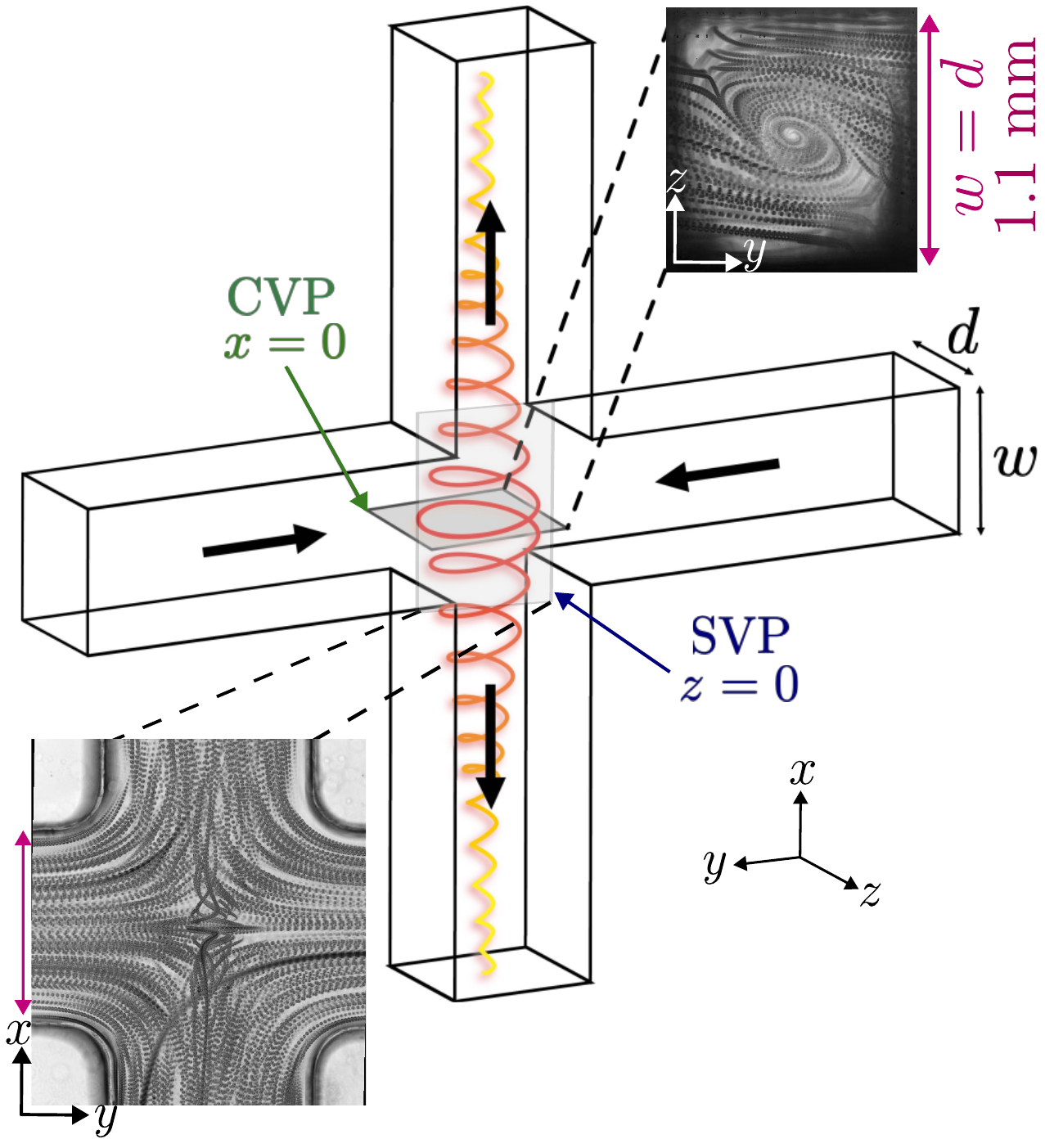}
\caption{Particle flow in the vortical field formed in the cross-slot geometry; Particle inflows from two opposing directions interact with a 3D vortex that is stretched downstream in the opposing outlet directions. Insets show superposition of experimental images of $20~\mu m$ particles in the flow ($Re = 56$) in both orthogonal planes of interest: the Core Vortex Plane (CVP) at $x = 0$ and the Stretched Vortex Plane (SVP) at $z = 0$.}
\label{fig:figure1}
\end{figure}

In this study, a microfluidic cross-slot geometry is used to induce a steady and stable three-dimensional (3D) vortical flow structure enabling investigation of particle dynamics. When inflows (along $y$-axis) from two opposite directions meet at the center of the geometry, a stagnation point is formed. A flow instability, in the form of a Burgers-like vortex stretched downstream (Figure~\ref{fig:figure1}), appears at a critical Reynolds number $\text{Re}_c \simeq 42$ for an aspect ratio $AR = w/d = 1$, where $w$ is the channel width and $d$ is the channel depth.
Here, we define the flow $\text{Re}$ as: 
\begin{equation}
    \text{Re} = \rho_f Uw/\eta,
\end{equation}
where $U = Q/wd$ is the average flow velocity in each one of the inlets and outlets, $Q$ is the volumetric flow rate (between 4 ml/min and 8 ml/min), $\eta$ is the dynamic viscosity, and $\rho_f$ is the fluid density~\cite{haward2016tricritical,burshtein2017inertioelastic}. The Reynolds number varies between $40 < \text{Re} < 80$, exploring the range where the vortex is stationary. Beyond these Reynolds numbers, a secondary instability appears: The vortex oscillates and eventually breaks down \cite{burshtein2021periodic}.

\subsubsection{Microfluidic devices}
In order to perform experimental visualizations in both planes of interest, two microfluidic devices are used: one made of glass for visualizations in the $x = 0$ plane, i.e. the Core Vortex Plane (CVP) and one made of PDMS for visualizations in the $z = 0$, i.e. the Stretched Vortex Plane (SVP). These devices have a square cross section $w=d=1.1$ mm, resulting in an $AR=1$. 

The glass microfluidic cross-slot device is fabricated with the selective laser-induced etching technique in fused silica with the ‘‘Lightfab’’ 3D printer (LightFab GmbH)~\cite{meineke2016,burshtein2019}. The cross-slot is embedded in one piece of glass and can withstand high pressure without deformation. To avoid unwanted flow instabilities, that may affect particle migration in the channel, the device is fabricated with long inlets ($19w$) and the injection ports are aligned with the channel inlets, allowing fully developed flow. The outlets are designed to be as long as possible ($5w$) while maintaining visibility of the CVP ($x=0$), a plane that is not accessible for imaging with standard devices~\cite{burshtein2019controlled,burshtein2019}. 

The PDMS device is fabricated using a brass mold of the geometry as the base of the device, then a classical lithography technique is used to fabricate the apparatus. While PDMS is not as rigid as glass, it can easily withstand the pressure we apply for our experiments. Here too, to avoid any flow instabilities, inlets and outlets are designed as long as possible. Note that this device gives us access to the more traditional plane of observation, the SVP plane, $z=0$.

To drive the flow in the channels, four 25 ml Hamilton Gastight syringes are individually controlled by high-precision neMESYS syringe pumps (Cetoni GmbH, Germany). Two of the pumps drive the suspension into the inlets, while the other two withdraw the suspension through the outlets, all pumps operate at an equal volumetric flow rate. Large syringes were used to allow the range of flow rates tested. Rigid poly(tetrafluoroethylene) (PTFE) tubing is used to connect between the syringes and the microfluidic device. The tubings are kept as short as possible in order to reduce system compliance to a minimum. The control over the flow field is done by adjusting the Reynolds number in the range $40 < \text{Re} < 80$.

\subsubsection{Suspensions}
Spherical polystyrene particles (Dynoseeds, MICROBEADS Norway) with density $\rho_p$ = 1050 kg~m$^{-3}$ are suspended in a 25 wt$\%$ glycerol (99$\%$ extra pure, Acros Organics) - DI water mixture ($\rho_f$ = 1059 kg~m$^{-3}$, $\eta$ = 1.79 mPa~s), resulting in almost neutrally buoyant particles. The concentration of the particles in the solution is determined according to their diameters $a$ in order to keep the number of particles roughly constant. For $a=$ 20, 40, and 80~$\mu$m, the resulting mass fraction in the suspension is $2\times10^{-4}$, $5 \times 10^{-4}$, and $1 \times 10^{-3}$, respectively. Therefore, suspensions are very dilute so that particle particle interactions can be neglected and particles are considered as single particles transported in the fluid flow. Additionally modifications of the base flow due to the presence of particles can be neglected. Particles with diameters larger than 10\% of the channel width ($a/w > 0.1$) were experimentally tested; however, in the working fluid used, they were not neutrally buoyant and exhibited sedimentation due to density mismatch.

To compare particle sizes and channel dimensions, we introduce the confinement ratio $\alpha = a / w$, which is in the range $0.018~\leq~\alpha~\leq~0.073$, and therefore considered low for this experimental system.

An additional parameter relevant to this study is the particle Stokes number, $St$, which compares the particle's kinetic energy to the energy dissipated by viscous diffusion in the fluid:
\begin{equation}
    \text{St} = \frac{1}{18} \frac{\rho_p}{\rho_f} \left( \frac{a}{w} \right)^2 \text{Re}.
\end{equation}
In our experiments, the particles are almost neutrally buoyant ($\rho_p / \rho_f \simeq 1$), resulting in a Stokes number in the range $0.001~\leq~\text{St}~\leq~0.02$. The Stokes number is typically used in particle-laden flows to characterize whether particles can be considered as passive tracers. In the specific case of density-matched particles, as considered here, the Stokes number becomes proportional to the particle Reynolds number:
\begin{equation}
    \text{Re}_p=\left( \frac{a}{w} \right)^2 \text{Re},
\end{equation}
which ranges from 0.02 to 0.35. We find that $Re_p<1$, consistent with conditions typical of inertial microfluidics, where particle migration is slow and particle trajectories deviate only slightly from streamlines on a local scale.

Note that, in our study, the particle Péclet number given by Pe $= 6 \pi \eta U a^2 / k_BT$, comparing thermal to advection effects, is in the range $0.3 \times 10^{9} < \textup{Pe} < 4.3 \times 10^{9}$, confirming that particles used are non-Brownian and that particle diffusion can be neglected.

\subsubsection{Data collection and analysis}
The microfluidic devices are mounted, one at a time, on a stage of a Zeiss microscope equipped with a 10x lens, with a depth of focus of 70 $\mu$m and numerical aperture $NA \simeq 0.2$. Focus is put either on the CVP, with the glass device imaged from the $yz$-plane, or the SVP, with the PDMS device imaged from the $xy$-plane. Each run of experiments is captured with a Photron fast camera at 2000 or 3000 fps for 1 second. The measurement depth, i.e., the depth over which we consider particles to be out of focus and no longer contribute significantly to the measurements, varies with the particle size \cite{meinhart2000}. It is given by:
\begin{equation}
    dz = \frac{3n\lambda}{(NA)^2} + 2.16 \frac{a}{tan\theta} + a,  
\end{equation}
where $n=1.364$ is the refractive index of glycerol 25$~\%$wt. in water, $\lambda = 550$~nm is the average wave length of the light source and $\theta=NA n_{air}^{-1}=0.2$~rad is the aperture half angle. The calculated measurement depth for which particles can be imaged is $dz=$ 398, 685, and 1258~$\mu$m for particles with a diameter $a = $ 20, 40, and 80~$\mu$m, respectively. The actual measurement depth is narrower according to the plane of interest we focus on. The procedure to reconstruct the particle trajectories is as follows: On each snapshot, the average background is removed, and a Gaussian filter is applied. A binarization step allows filtering the particles that are out of plane (the threshold for binarization varies for the CVP and SVP), and the coordinates of the remaining particles are extracted. Finally, particle trajectories are calculated using a nearest-neighbor algorithm. For experiments of the CVP, the binarization threshold is high so that the measurement depth is of the order of one particle diameter; particles not located exactly in the plane are filtered out. However, for experiments focusing on the SVP, in order to follow the particles in a quasi-3D manner, the binarization threshold is lower such that particles located several particle diameters above or below the SVP are kept for subsequent analysis.

\begin{figure}[t]
\centering
\includegraphics[width=14.25cm]{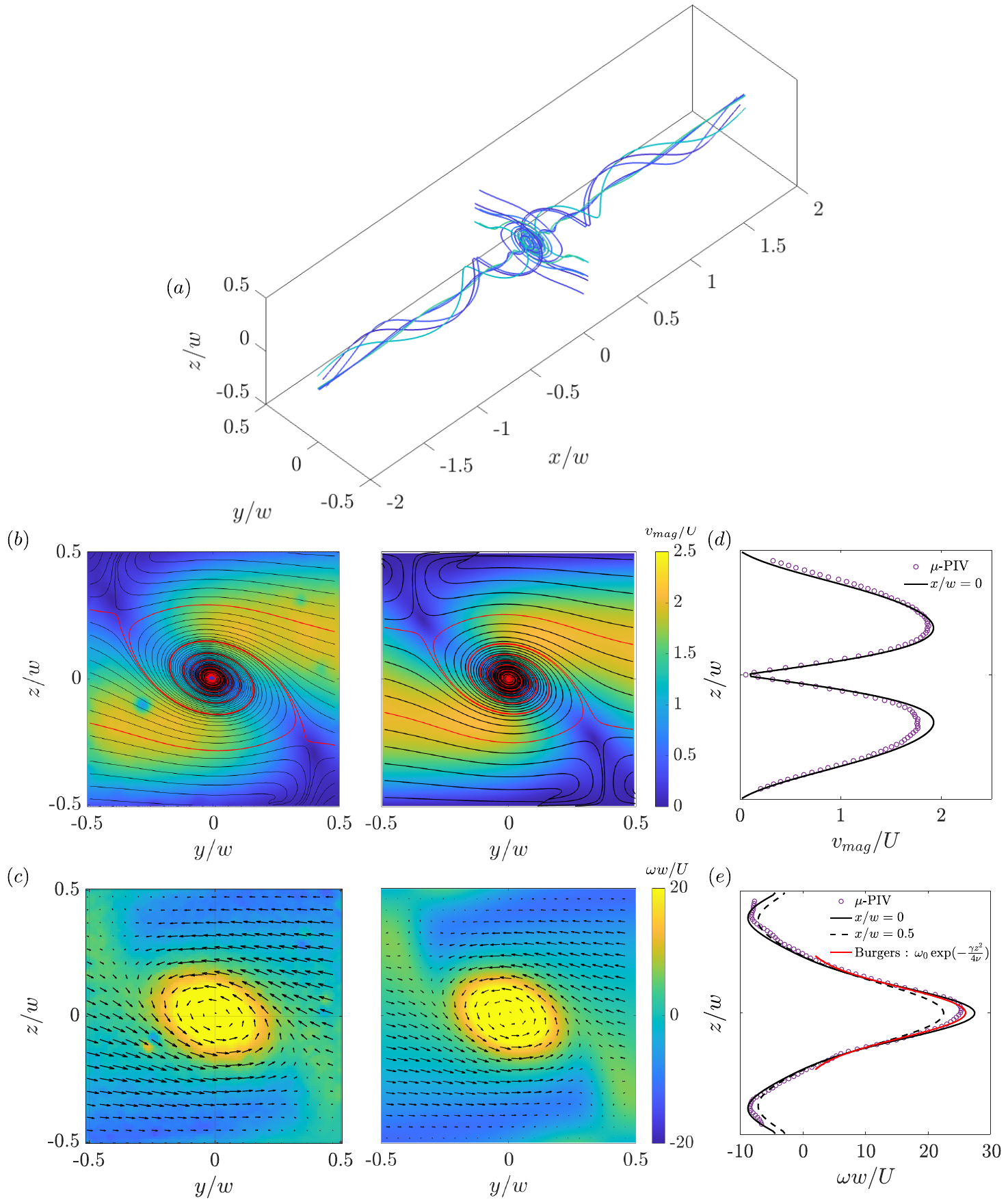}
\caption{Base vortical flow field at $Re = 56$. (a) Simulated 3D streamlines of the base flow wrap and spiral around the vortex core. (b) Streamlines superimposed to the normalized velocity magnitude of the base flow from experiments (left) and simulations (right). Red lines mark the border of the streamlines interacting with the central vortex. (c) Normalized vorticity field of the base flow from experiments (left) and simulations (right). (d) Dimensionless velocity profiles along $z/w$ at $y/w=0$ and $x/w=0$. (e) Dimensionless vorticity profiles along $z/w$ at $y/w=0$ and $x/w=0$ and $x/w = 0.5$. The core vorticity is also fitted with the profile $\omega(z) = \omega_0 \exp(-\gamma z^2 / 4\nu)$ (red line, with fitting parameters $\omega_0 = 28$, $\gamma/4 \nu = 0.01$), matching the vorticity profile of a Burgers vortex.}
\label{fig:figure2}
\end{figure}

\subsection{Numerical method}
Direct numerical simulations of the flow and particle dynamics were performed for $Re = 56$ and two particle sizes, $a = 80$ and 140 $\mu$m, to complement experimental measurements and access another confinement value $\alpha = 0.127$. For both sizes, simulations are performed with respectively 20 and 14 different initial positions, while assuring distances between particles are sufficiently large to avoid particle particle interactions. The extremely small particle concentration also ensures that changes to the base flow can be neglected after time averaging. For simulation of the fluid phase, we use SNaC, a robust multi-block finite-difference solver~\cite{costa2022fft}. The fluid and solid motion are coupled by means of the immersed-boundary method~\cite{breugem2012second}. In brief, we adopt two different grids: one for the fluid flow, which is a stationary Eulerian grid, and a second, moving grid attached to the particles (Lagrangian grid).
Concerning the fluid-structure interaction, first, the flow velocity is interpolated on the Lagrangian grid and used to compute hydrodynamic forces. These hydrodynamic forces are then used to update the position and velocity of each particle. At the last step, the hydrodynamic force is spread back to the Eulerian grid and used as a source term in the Navier-Stokes equations to model the presence of the solid object when updating the flow field. The method has been utilized in several studies \cite{breugem2012second,picano2015turbulent,lashgari2015transition,fornari2016sedimentation,lashgari2016channel,ardekani2016numerical,ardekani2018heat,yousefi2023role}, providing accurate results. For this work, we have implemented the same immersed-boundary method in SNaC with some additional parallelization considerations for performance improvements. The new code has been validated against several benchmark cases, and the results are in perfect agreement with previous validations against experimental and theoretical results.\\
The computational domain is designed to reproduce the experimental setup, consisting of a cross-shaped channel with two inflows and two outflows. The inlet and outlets have a square cross section with side $w$ equal to 1.12 millimeters. The inlet and outlet ducts have lengths equal to 8.96 millimeters $(8w)$. Uniform velocity and zero-gradient velocity boundary conditions are imposed at the inlets and outlets. 
The computational grid is uniform throughout the domain. The grid resolution is chosen to capture the flow around the particles, corresponding to at least 16 grid points per particle diameter for the moderate $\text{Re}$ considered in this study. Specifically, for the simulation of the large particles, $140~\mu$m in diameter, we use 20 grid points per particle diameter, whereas 16 grid points per diameter are utilized for the $80~\mu$m particles. This results in $160$ and $224$ grid points along the channel side for the $140~\mu$m and $80~\mu$m particles, respectively. Smaller particle sizes are not accessible because the simulations would become too costly. Hence, 80 $\mu$m particles are the only particle diameter in common with the experiments.
Before injecting the particles into the computational domain, the background flow field is generated. In order to capture the vortex region at the cross-junction, the flow field is initialized with zero velocity except in the middle of the channel, where some noise is introduced to break the symmetry; the solution is advanced in time until it reaches a steady state. The resulting flow field is then used as an initial condition for the simulations with particles. The particles are injected around the $x=0$ (CVP), some exactly at the CVP, some above and below it (demonstrated in Section \ref{sec:particle_dyn} below), at a distance $\pm a/2$, where $a$ indicates the particle diameter. The initial translational and rotational velocity of the particles are set to zero.

\subsection{Base flow characterization} \label{sec:PIV}

Quantitative measurements of the base flow, in the 25$\%$ w/w aqueous glycerol solution, are performed in the CVP using micro-particle imaging velocimetry ($\mu$-PIV). The flow is seeded with fluorescent particles with a diameter of $3.2~\mu$m and excitation and emission wavelengths of 542 nm and 612 nm, respectively. The glass cross-slot device is mounted on the stage of the Zeiss inverted Microscope, and the focus is set on the $x = 0$ plane (CVP). The $\mu-$PIV system is equipped with a LaVision set-up composed of a Nd: YAG double pulsed Laser and a double framed CX-Imager Camera, both controlled via a Programmable Time Unit. During each pulse, the fluid and the tracers are illuminated with a 520 nm wavelength, thus exciting the particles, and then they emit at a longer wavelength. The camera filters and captures the reflected light. Each velocity field is calculated between two pairs of image frames with Davis Software. Base flow characterization is focused only on the CVP. Performing measurements in the orthogonal SVP will not provide additional information on the base flow since streamlines move in and out of the plane, and cannot be captured with 2D imaging.

Figure~\ref{fig:figure2} shows a comparison between simulations and $\mu-$PIV measurements of the base flow for $Re = 56$. A global view of the flow is shown in panel (a) with a 3D plot of streamlines extracted from numerical simulations. Streamlines starting close to the CVP ($x = 0$) will wrap more tightly around the vortex core than streamlines starting further away. The velocity and vorticity fields are shown in panels (b) and (c), respectively. The inflows meet at the center of the geometry to form the 3D vortex.  In the CVP, specific streamlines, hereafter called separatrix streamlines (highlighted in red in panels (b) and (c)), separate different regions of the flow. Outside of them, streamlines do not directly interact with the stagnation point, while inside of them, they will wrap around it. Another key feature of the base flow is the highly concentrated region of vorticity around the stagnation point. Velocity and vorticity fields from simulations and experiments are in good agreement. This agreement is further confirmed with panels (d) and (e) showing velocity and vorticity profiles along $z/w$ at $y/w = 0$ in simulations and experiments. The vorticity profiles can be extracted from the simulations for different $x$ planes and reveal that the concentrated vorticity region remains along the $x$ axis. However, it weakens farther from the mid-plane towards the outlets. The size of the vorticity region moderately changes in the cross-slot area. This flow structure resembles the Burgers vortex: vorticity is concentrated in a narrow axisymmetric column (centered at the stagnation point) due to the balance between flow stretching and viscous diffusion. Note that the core vorticity profile closely follows a Gaussian distribution. This is consistent with the vorticity profile of a Burgers vortex. Figure~\ref{fig:figure2} (e) shows this agreement based on the analytical expression derived from the orthoradial velocity  $u_\theta(z)=\frac{2\nu\omega_0}{\gamma z} \left(1-\exp\left(-\frac{\gamma z^2}{4\nu}\right) \right)$ where $\gamma$ is the strain rate, $\nu$ is the kinematic viscosity and $\omega_0$ is the vorticity at the core center.

\begin{figure}[t]
\begin{center}
\includegraphics[width=13cm]{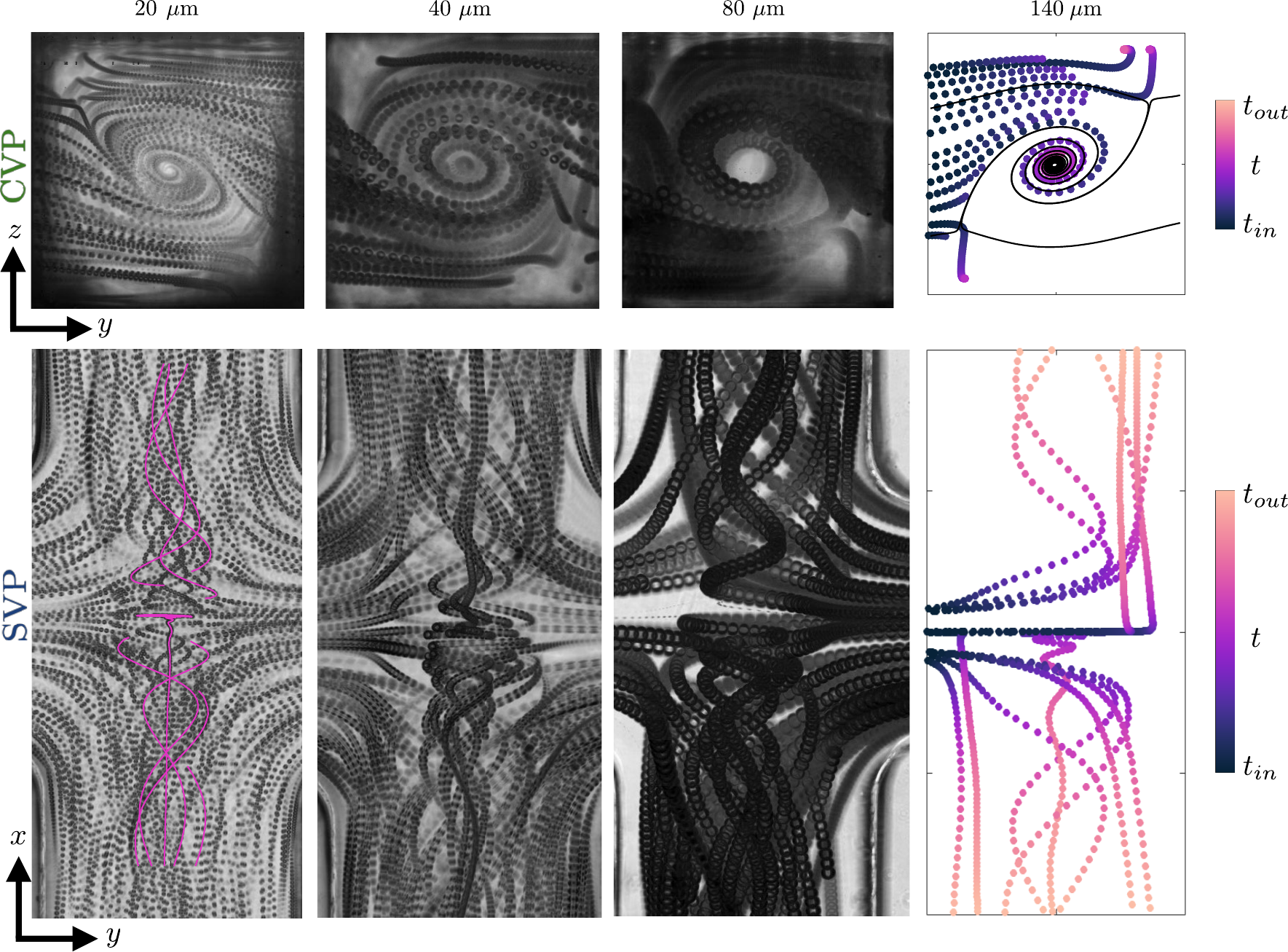}
\caption{Stacks of experimental and simulated particle images in the cross-slot geometry at $Re = 56$.  Particle diameters are 20, 40 and 80~$\mu$m for experiments and 140 $\mu$m for simulations. Top row: visualization of the Core Vortex Plane (CVP), obtained with a microfluidic glass device. Bottom row: Stretched Vortex Plane (SVP), imaged using a PDMS device. Examples of the reconstruction of particle trajectories are obtained from experimental data and overlaid as magenta lines in the 20~$\mu$m panel. Trajectories obtained from simulations for 140 microns particles are color-coded with time with $t_{in}$ the entry time of the particles and $t_{out}$ the exit time.}
\label{fig:figure3}
\end{center}
\end{figure}

\section{Particle dynamics} \label{sec:particle_dyn}

Here, we aim to understand the behavior of spherical particles in a vortex flow and study the effects of $\text{Re}$ and particle size. For this purpose, we focus on two observation planes: the CVP to conduct a statistical study on the number of trajectories in the plane and the SVP to closely follow the particles and the flow along each individual trajectory. Figure~\ref{fig:figure3} shows the superposition of experimental images in both planes for three sizes of particles. Experiments are conducted in a dilute suspension: For each individual snapshot, only a couple of particles are visible in the field, allowing us to track them easily in time and obtain their trajectories.

\begin{figure} [t]
    \centering
    \includegraphics[width=10.5cm]{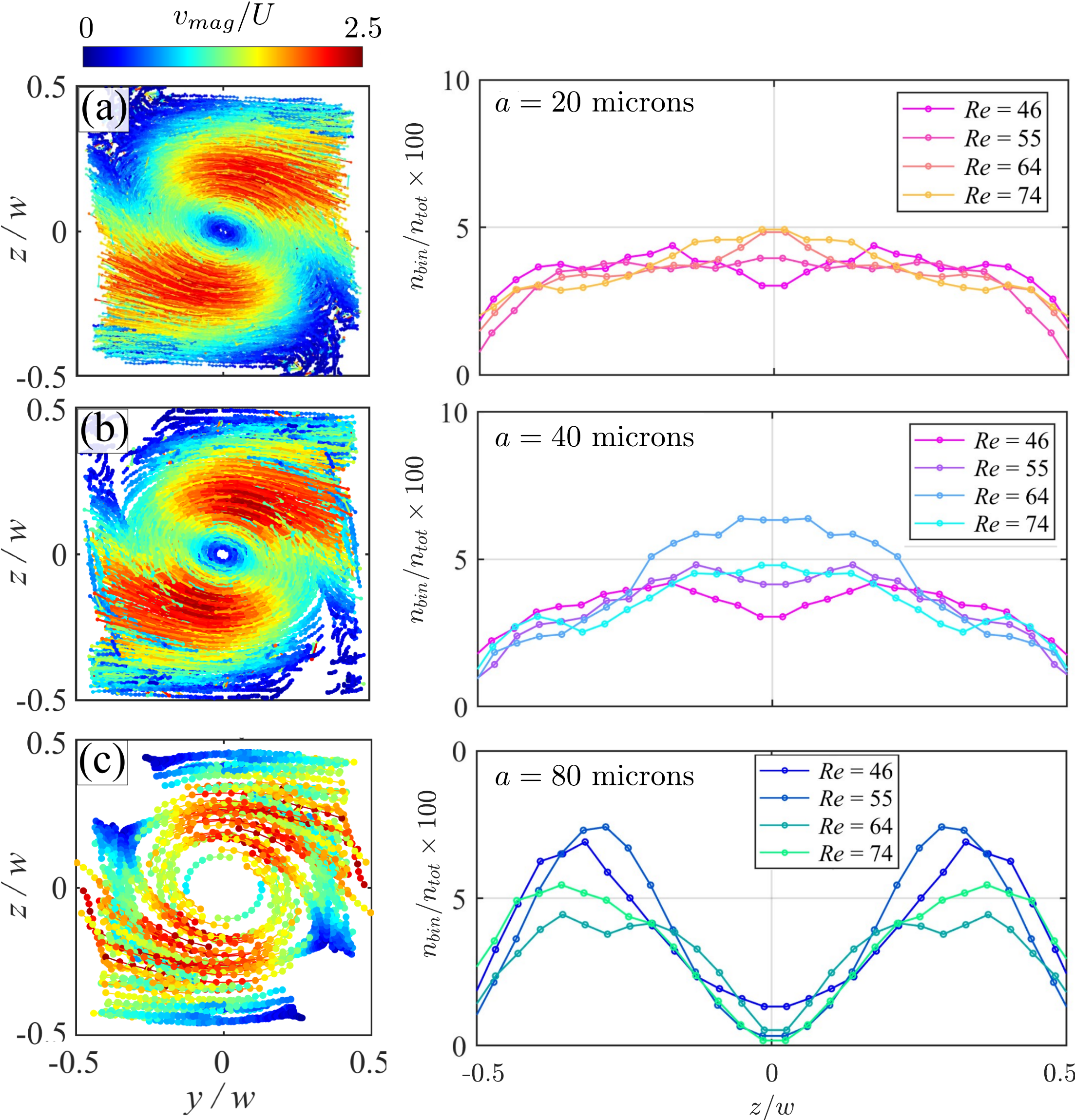}
    \caption{Effect of particle size on trajectories and density distribution in the CVP. Left panels: Particle trajectories extracted from experiments in the $x=0$ (CVP) plane, color-coded by normalized velocity magnitude. Experiments were conducted at $Re=56$ for particle diameters: (a) $a=20$ $\mu$m; (b) $a=40$ $\mu$m; and (c) $a=80$ $\mu$m. Data are averaged and mirrored around the axis $z = -y$. Right panels: Corresponding normalized particle density profiles along $z/w$ measured at $x/w=y/w=0$. Particle density is computed as $n_{bin}/n_{tot}*100$ where $n_{bin}$ is the number of particles within each bin (bin width = 0.04) and $n_{tot}$, is the total number of particles in the observed plane.} 
    \label{fig:figure4}
\end{figure}

This stack representation gives us a clear overview of several characteristics of the particle dynamics. Two distinct types of trajectories emerge: some interact with the vortex core, spiraling around it while being advected downstream, while others avoid it and are simply advected towards the outlets. These observed behaviors are in agreement with trajectories obtained from numerical simulations (see 140 microns panel in Figure~\ref{fig:figure3}). Additionally, it is visible from Figure~\ref{fig:figure3} that particle size influences 
dynamics: larger particles are less likely to reach the vortex core. This observation is confirmed by simulations for 140 microns particles, which show that large particles leave the CVP early (at $x = 0$) before reaching the vortex core. In the following sections, we propose a more detailed analysis of these phenomena. 


\subsection{Statistical study} \label{sec:statistics}

Before conducting a systematic study of individual trajectories, we first provide a broader view of the problem, focusing on the CVP ($x = 0$). Figure~\ref{fig:figure4} presents a superimposition of particle trajectories obtained experimentally for three particle sizes between 20, 40, and 80 $\mu$m. Around 100 trajectories are superimposed for each particle size. The trajectories are color-coded by normalized particle velocity magnitude. 
The velocity distribution closely resembles that of the base flow (Figure~\ref{fig:figure2}). In all cases, particles exhibit higher velocities upon exiting the inlets and just before circling the vortex core. This behavior appears consistent across all particle sizes tested. Due to spatial variations in velocity across the flow field, particles experience 
different residence times within the measurement frame. 

Secondly, the particle density along $z/w$ is measured experimentally for the 3 particle sizes by counting how many particles cross the $y/w = 0$ line (Figure~\ref{fig:figure4}, right panels), across 4 different Re.
We notice that no significant differences are observed as a function of Re, within the range of Reynolds numbers tested. However, particle size has a clear effect; as size increases, particles are less likely to enter the vortex core. 
If the particles behaved as passive tracers, we would expect them to be drawn toward the stagnation point at the vortex center. The closer a particle gets to the stagnation point, the longer it should remain near it. Therefore, a peak in the density distribution at the vortex core could be expected, reflecting a theoretical infinite residence time. However, the observed distribution shows the opposite: larger particles avoid the vortex center. 
This suggests that finite-size and/or fluid inertia alter particle trajectories and prevent them from reaching the vortex core. In reality, a finite-size particle experiences multiple streamlines and is influenced by the full complexity of the 3D flow structure. This observation motivated the need to shift our focus on 3D particle trajectories, as available from numerical simulations, to capture the full 3D dynamics of particles within the vortex.

\begin{figure}[p]
\centering
\includegraphics[width=14.75cm]{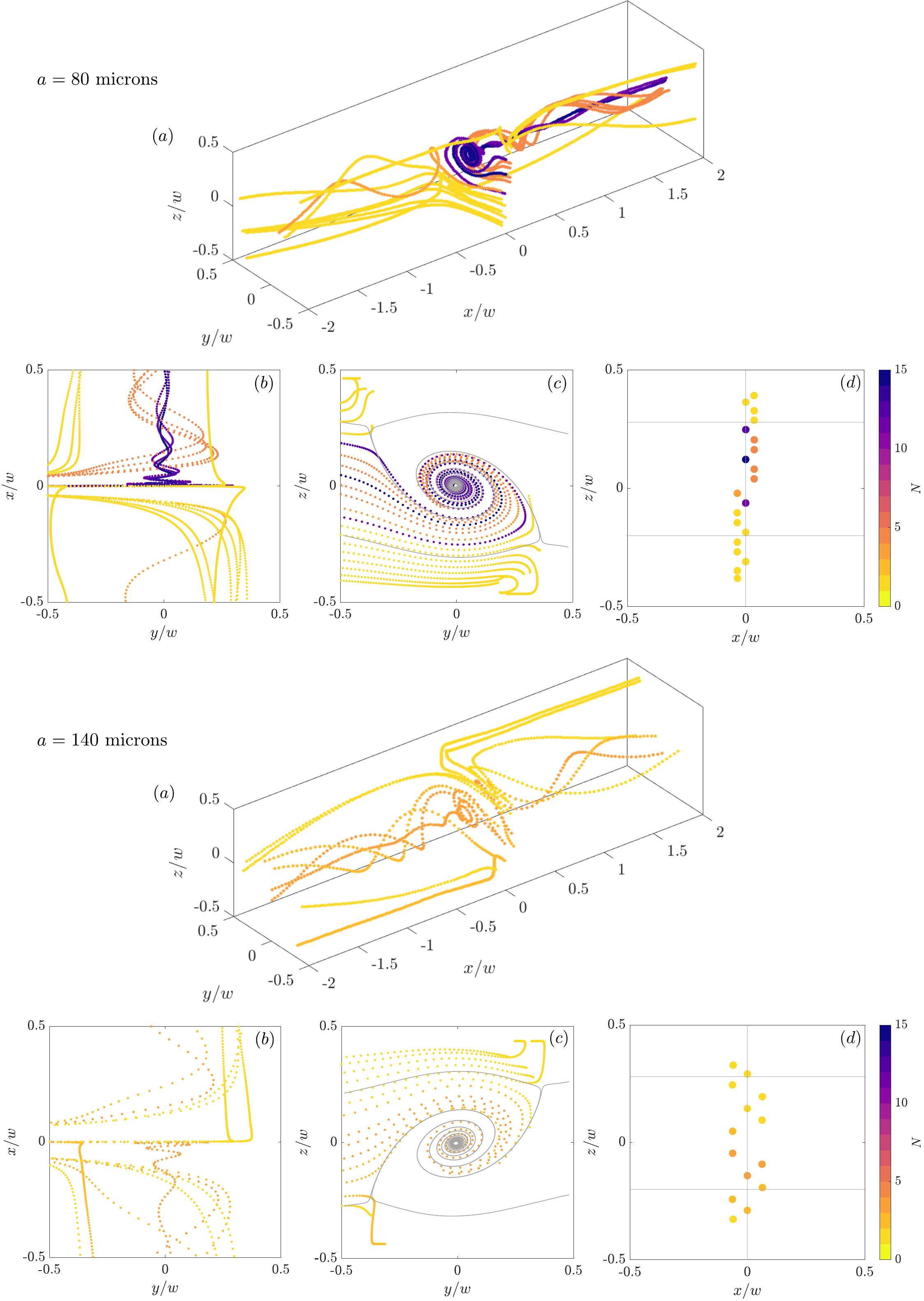}
\caption{Types of particle trajectories from simulations for 80 $\mu$m particles (top) and 140 $\mu$m (bottom)  at $Re = 56$. Two distinct behaviors are observed: particles that spiral around the stagnation point while being advected into the outlets and particles that are advected without interacting with the vortex core. All panels are color-coded based on the number of spirals \textit{N} a particle completes around the vortex axis. (a) 3D Visualization of particle trajectories in the cross-slot and then the outlets; Particle trajectories projection: (b) in the SVP; and (c) in the CVP (volume $|x| < 0.5$). (d) Initial positions of particle trajectories (lines indicate separatix streamlines at the cross-slot entry)}
\label{fig:figure5}
\end{figure}

\subsection{Study of typical trajectories}
We now examine individual particle trajectories based on numerical data, as shown in Figure~\ref{fig:figure5} (panels a, b, and c) for 80 and 140 $\mu$m particles, respectively. Two distinct types of trajectories emerge: (i) those that spiral around the vortex core  (shown in darker shade), and (ii) those that are advected away from the core and remain largely unaffected by it (lighter shade). 
For both types of trajectories, the corresponding initial particle positions are plotted in panel (d) of the same Figures with the same color code. At first glance, observations show that particles injected close to the CVP are more likely to follow spiral trajectories around the vortex.

\begin{figure}[t]
\centering
\includegraphics[width=15cm]{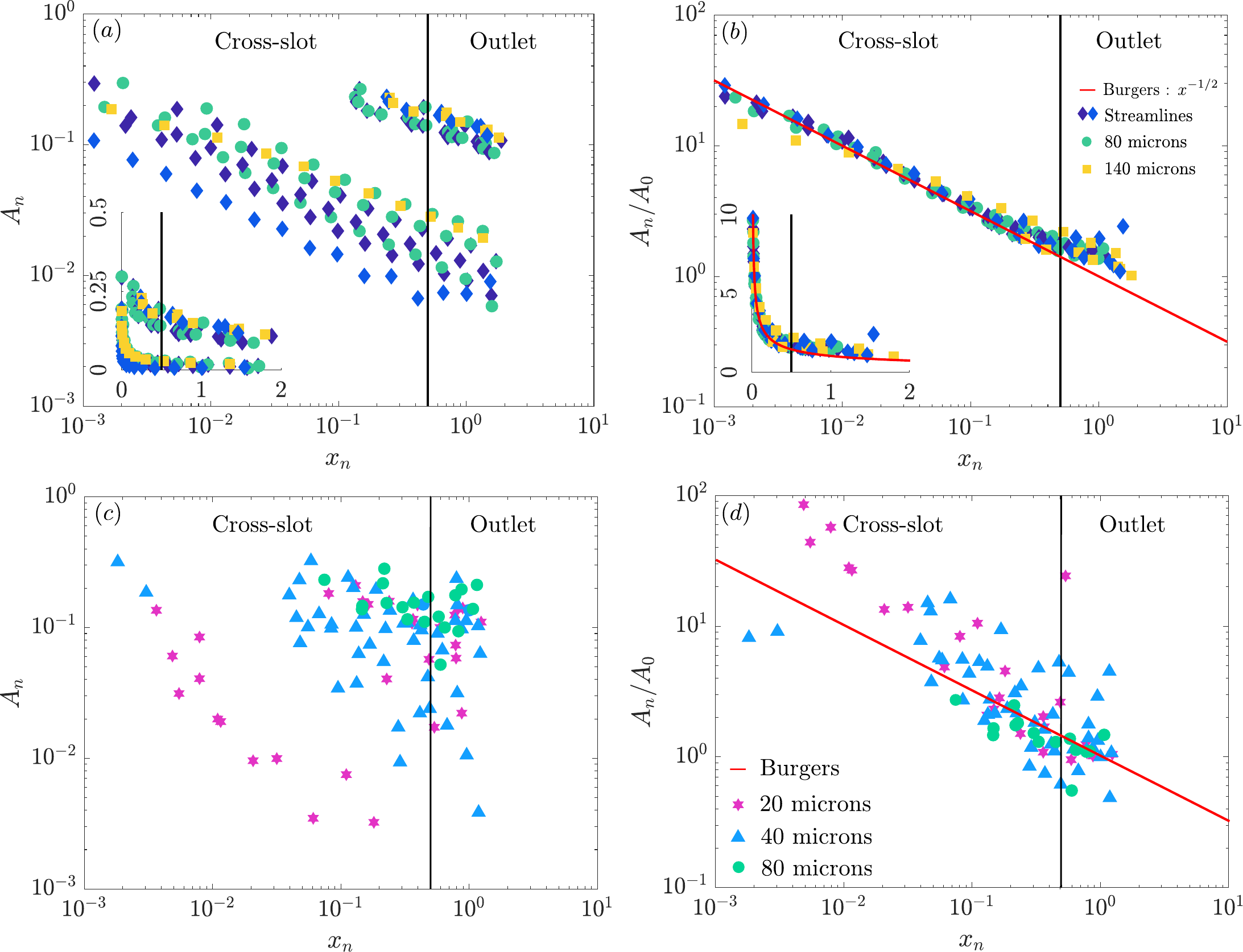}
\caption{Decay of spiral amplitude along the vortex axis. (a) Amplitude of the spirals $A_n$ as a function of their position $x_n$ on a log-log scale. Diamond markers represent streamlines. \textcolor{Blue}{$\blacklozenge$} represent streamlines starting at the same initial positions of the 80 $\mu$m particles, \textcolor{NavyBlue}{$\blacklozenge$} represent streamlines with the same initial positions as the 140 $\mu$m particles. \textcolor{SeaGreen}{\Large\textbullet} represent 80 $\mu$m particles and \textcolor{Goldenrod}{$\blacksquare$} represent 140 $\mu$m. The inset shows the linear representation of the same data. (b) Normalized amplitude of spirals, $A_n / A_0$ as a function of position $x_n$ in log-log scale, compared with a decaying envelope of the Burgers vortex streamlines (red solid line). The normalization coefficient $A_0$ is the fitting coefficient of each power law $A_n = A_0 x_n ^{-0.5}$. The inset presents the linear representation of the same plot. (c) Experimental amplitude $A_n$ of spirals as a function of their position $x_n$. (\textcolor{VioletRed}{$\bigstar$} 20 $\mu$m, \textcolor{Cyan}{$\blacktriangle$} 40 $\mu$m and \textcolor{SeaGreen}{\Large\textbullet} 80 $\mu$m).  (d) Normalized experimental amplitude $A_n / A_0$ as a function of their position $x_n$.}
\label{fig:figure6}
\end{figure}

To quantify the extent of a particle's interaction with the vortex, we follow the methodology as in~\cite{kechagidis2023numerical}. Specifically, we measure the number of spirals, $N$, a particle completes around the vortex axis, by identifying each local maximum in its trajectory and mapping its position in the SVP (Figure~\ref{fig:figure5}, panel (a)). Panel (d) of Figure~\ref{fig:figure5} (for 80 and 140 microns) shows the particles' initial positions, color-coded by their corresponding spiral count $N$. These results show that the type of trajectory a particle follows, whether it spirals around the vortex or bypasses it, is determined primarily by its initial position. Notably, the initial $y$ position has no influence on the path followed by a particle. 
In the experiments, particles are injected at the start of the inlets. However, in simulations, particles are carefully ``released'' upstream of the cross-slot region so they are at equilibrium with the flow before entering the vortex formation region. The initial $z$-position, i.e., the vertical location at which a particle enters the CVP, plays an important role. If a particle enters the cross slot region between the separatrix streamlines, it is more likely to interact with the vortex and spiral around it, corresponding to a higher $N$. 
However, if a particle enters outside the separatrix streamlines, it interacts primarily with the corner vortices and is diverted away from the main vortex, eventually being advected directly into the outlets. Finally, the initial $x$ position, i.e., the particle's distance from the CVP, also determines the number of spirals it performs around the vortex core. Particles initialized closer to the CVP are more likely to rotate around the vortex axis before being advected out through the outlets. 
This can be explained by the flow's symmetry with respect to the CVP: a particle entering exactly at $x = 0$ experiences equal flow magnitude on each side of the midplane, making it more likely to swirl around the vortex axis. In contrast, for particles entering farther from the midplane are influenced by the 3D flow structure and are pushed outward, resulting in fewer spirals around the vortex axis.
A comparison between top and bottom panels of Figure~\ref{fig:figure5} illustrates this effect, showing that the larger particles generally spiral less around the vortex core (particularly in panel (d) of both Figurs). Most of the 140 $\mu$m particles are injected further away from the CVP ($x = 0$) than the 80  $\mu$m particles and, therefore, are directly advected to the outlets. However, for those who are initialized at the CVP, the 140 $\mu$m particles complete fewer loops around the vortex axis ($N_{max,140} = 7$), while transported over the same downstream distance, than the 80 $\mu$m particles ($N_{max,80} = 15$). This suggests that, for larger particles, the advective motion towards the outlet becomes more important compared to the rotative motion around the vortex. This is in agreement with the statistical observations; bigger particles tend to interact less with the vortex core.

Next, we examine the trajectories that spiral around the vortex core. These correspond to initial positions where $x$ is close to the CVP and $z$ lies between the separatrix streamlines within regions of high $N$). In Figure~\ref{fig:figure6} (a) and (c), the amplitude of each spiral around the vortex core, denoted $A_n$ (i.e. particle radial distance to the vortex rotation axis), is measured and plotted as a function of its downstream distance, $x_n$, for both experimental and numerical datasets, respectively. 
Two observations can be made. 
First, the radius of the first loop is roughly constant for all initial positions for both numerical and experimental cases tested as we can see in Figure~\ref{fig:figure6} (a). As shown by the PIV measurements discussed in section~\ref{sec:PIV}, the high-vorticity region around the stagnation point does not vary significantly in the cross-slot. Therefore, the first interaction of the particle with the vortex always occurs at the same distance from the core. As noted in section~\ref{sec:PIV}, this flow is reminiscent of a Burgers vortex, in which a balance between viscous diffusion and vortex stretching leads to a concentration of vorticity along a cylindrical region around the rotation axis, with a Gaussian distribution. 
Secondly, the amplitude of the spirals decreases as the particle is advected to the outlets. The particle spiral motion attenuates following a decreasing power law, $A_n = A_0 x^n$. Upon normalizing by the fitting coefficient $A_0$ for each trajectory, we see that the numerical data collapse on a master curve with a main slope of $n = -1/2$. The attenuation is slightly more pronounced (i.e., with a steeper slope) for small values of $x_n$, corresponding to regions near the CVP, than for larger values of $x_n$, located further downstream outside of the cross-slot where the vortex is more confined. This trend is observed using the streamlines (diamond markers) and is superimposed on the particle trajectory data (circle and square markers) from the numerical simulations (top panels of Figure~\ref{fig:figure6}). 
For 80 microns particle, the deviation from the streamlines is negligible. However, the decay rate of the 140 $\mu$m particles that reach the vortex core is slower than that of the base flow, with an approximate scaling of $A_n \simeq A_0 x^{-0.35}$. This means that all particles may eventually reach a similar minimum radial distance from the core, however larger particles will reach it further downstream.

To understand this reduced decay rate, we return to the comparison with the axisymmetric Burgers vortex, in which the velocity along the rotation axis is defined as $u_x = \gamma x$ where $\gamma$ is the stretching rate. The incompressibility condition of the flow yields the following expression for the radial velocity component $u_r = - \frac{\gamma}{2} r$. 
Since the flow is axisymmetric, the azimuthal component of the velocity is neglected. We, therefore, consider a 2D configuration and calculate the trajectories of particle tracers as follows:

\begin{equation}
    \left\{
    \begin{array}{ll}
         \frac{d x}{dt} = u_x = \gamma x
         \\
        \\
        \frac{d r}{dt} = u_r = - \frac{\gamma}{2} r.
    \end{array}
\right.
\end{equation}

From this, it can be shown that the envelope of the trajectories satisfies a power-law with an exponent $-1/2$ :

\begin{equation}
     r = x^{-1/2}.
     \label{eq:eq_burgers}
\end{equation}

Therefore, in a typical Burger's vortex, the streamlines envelope decays along the rotation axis following a power law with an exponent of $-1/2$. Our simulations are mostly in excellent agreement with this law. This shows that particles up to 80 $\mu$m behave almost as passive tracers within a Burgers vortex. However, the 140 $\mu$m particles deviate slightly from the power law, with a more gradual decay in their spiraling motion. We also note that all particles and streamlines in our configuration deviate from the Burgers vortex envelope upon exiting the cross-slot region and entering the outlets, meaning that confinement from the wall slightly affects the fluid flow.    
Finally, the same procedure is applied to the experimental data set obtained from the 3D trajectory reconstruction in the SVP, shown in the bottom panels of Figure~\ref{fig:figure6}. Applying the same power law fit as for the numerical simulation reveals once again that all the data collapse on a master curve as shown in panel (d). This confirms that particles exhibit a decaying self-similar motion within the vortex. This main trend is in reasonable agreement with the Burgers vortex envelope (equation~\ref{eq:eq_burgers}). Note that in the experiments, it is more to difficult to control the baseflow precisely and to accurately position the observation plane within the SVP because particles enter and exit the plane periodically, which may account for the broader scatter observed. 

In conclusion, the spiraling motion of particles decays according to a power law as they are advected downstream, exhibiting a self-similar behavior. This motion is the result of the base flow dynamics governed by the balance between viscous diffusion and vortex stretching, analogous to the Burgers vortex. However, larger particles exhibit a slight deviation from the base flow, with a slower attenuation of their motion compared to smaller particles. This indicates that larger particles tend to be displaced from the vortex core. These observations suggest that particle size or fluid inertia plays a significant role in modulating particle-vortex interactions.

\subsection{Discussion}

\begin{figure}[t]
\centering
\includegraphics[width=14cm]{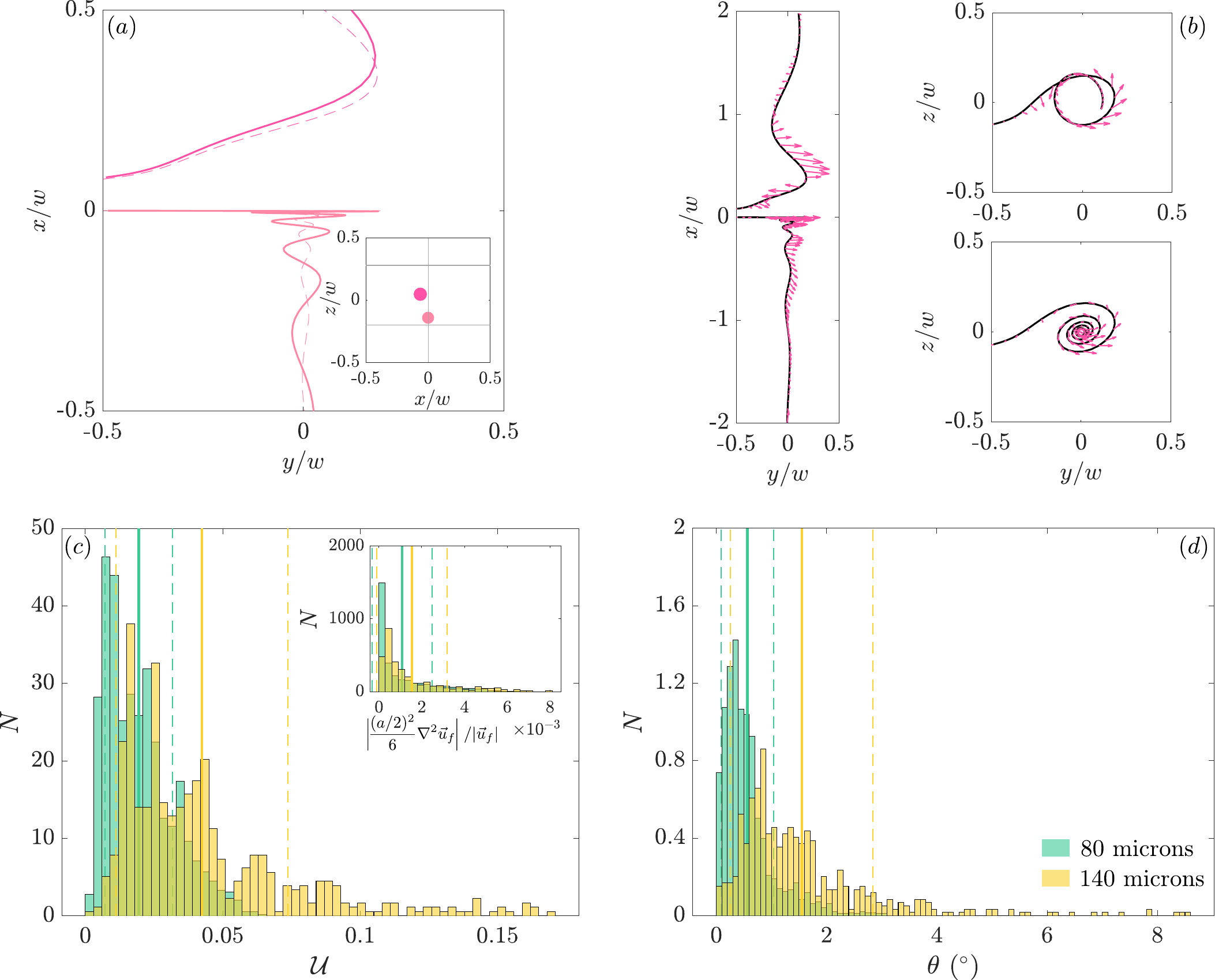}
\caption{Slip Velocity and Trajectory Deviation of Inertial Particles. (a) Comparison between two typical trajectories of 140 $\mu$m particles (thick line) and their corresponding streamlines (thin line) initialized from the same initial position. Initial positions are shown in the inset. (b) Projection of fluid and particle velocity difference (dark pink arrows) along the 140 $\mu$m particle trajectories on the CVP and SVP. Statistical analysis of slip velocity and alignment for 80 $\mu$m and 140 $\mu$m particles (c) PDF distribution of the normalized velocity difference between particles and fluids defined as $\mathcal{U} = | \vec{u_p} - \vec{u_f}| / |\vec{u_f}|$, over the entire domain. The inset shows the Fax\'en contribution to the velocity difference;  (d) PDF of the angle $\theta = acos \left( (\vec{u_p} \cdot \vec{u_f}) / |\vec{u_p}| |\vec{u_f}| \right) \times 180 / \pi$ between the particle and fluid velocities, quantifying their alignment or deviation. Solid lines represent the average value of each distribution, while dashed lines represent the standard deviation from the average.
}
\label{fig:figure7}
\end{figure}

As particle size is increased, particles are more likely to depart from the single phase base flow (i.e. the vortex flow without particles) and no longer follow the streamlines. Figure~\ref{fig:figure7} a shows two trajectories (solid lines) and their associated streamlines (dashed lines), starting from the same initial positions. Although the deviations are not large, they are sufficiently pronounced to observe that particles are displaced outwards from the streamlines and from the vortex as they are drawn towards the outlets. These deviations become more significant as particle size is increased and for higher streamline curvature. Through the simulations, we have access to the baseflow fluid velocity field along each particle trajectory. By examining the slip velocity between the fluid and the particles, projected onto both CVP and SVP (dark pink arrows), we note that particles are indeed pushed away from the vortex core. This is particularly evident where the trajectories spiral, as shown in Figure~\ref{fig:figure7} (b) and in agreement with our statistical trends shown in Figure~\ref{fig:figure4}. 

Several factors may account for the observed deviation from the base flow. It is important to note that the particles are practically neutrally buoyant; therefore, centrifugal forces cannot be responsible for the deviation. However, fluid inertia ($Re = 56$), flow curvature, and finite size effects may all contribute to the observed behavior. To investigate this further, we consider the normalized velocity difference $\mathcal{U}$ and the angle $\theta$  between the fluid and particle velocities, defined respectively as:
\begin{equation}
    \mathcal{U} = \frac{|\vec{u_p} - \vec{u_f}|}{|\vec{u_f}|} \quad \textup{and} \quad \theta = \arccos\left( \frac{(\vec{u_p} \cdot \vec{u_f})}{|\vec{u_p}||\vec{u_f}|} \right) \times \frac{180}{\pi},
    \label{eq:vel_diff}
\end{equation}

with $\vec{u}_p$ the particle velocity and $\vec{u}_f$ the fluid velocity at the particle center of mass position in the single phase baseflow.

\begin{figure}[t]
\centering
\includegraphics[width=9cm]{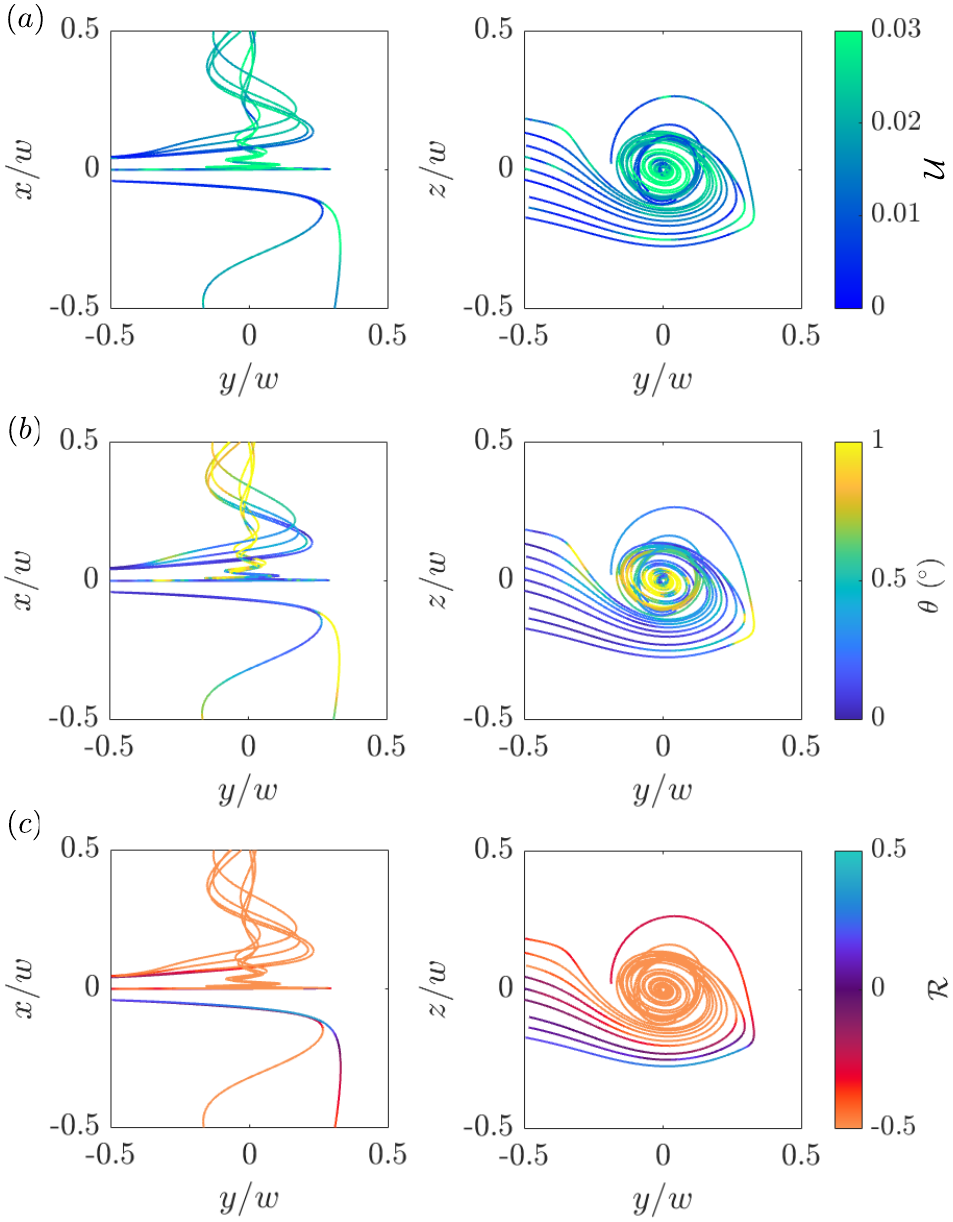}
\caption{Spatial distribution of particle–fluid discrepancies for 80 $\mu$m particles. Each quantity is shown along particle trajectories, color-coded in both the SVP (left panels) and the CVP (right panels). (a) Normalized velocity difference $\mathcal{U}$; (b) Angle $\theta$ between particle and fluid velocities; and (c) Rotation difference projected onto the vortex axis $\mathcal{R}$}. 
\label{fig:figure8}
\end{figure}

Figure~\ref{fig:figure7}, panels (c) and (d), show the probability density functions (PDFs) of the normalized velocity difference $\mathcal{U}$ and angle $\theta$, respectively, for 80 $\mu$m particles (green) and 140 $\mu$m particles (yellow). Each statistical distribution is positively skewed with a long tail toward higher values of the corresponding quantity. To provide a simplified quantitative interpretation, we calculate the average and standard deviation for both the velocity difference and deviation angle. This yields $\mathcal{U}_{80} = 0.02 \pm 0.005$ and $\mathcal{U}_{140} = 0.04 \pm 0.03$ for 80 $\mu$m and 140 $\mu$m particles, respectively. We conclude that $\mathcal{U}$ increases with particle size, but remains relatively small with approximately 2\% and 4\% of the base flow velocity. For the deviation angle, we obtain $\theta_{80} = 0.8 \pm 0.05 ^{\circ}$ and $\theta_{140} = 1.8 \pm 1 ^{\circ}$. The average angle between the particle and fluid velocity vectors increases with particle size, indicating that the velocities become progressively less aligned. This misalignment implies that larger particles will migrate away from the single-phase flow streamlines. Note that both distributions become broader for larger particles. Spatial maps of the normalized velocity difference and deviation angle (Figure~\ref{fig:figure8} (a-b)) show that the most pronounced differences from the base flow occur around the vortex core. These are particularly evident for strongly curved trajectories that spiral around the core. This deviation coincides with the regions of strongest misalignment between particle and fluid velocities.

To understand the differences between particle trajectories and flow streamlines, we first investigate finite-size effects, which persist even in the limit of vanishing Reynolds number. These effects are captured by the Fax\'en laws, which describe corrections to the translational and rotational motion of finite-size particles: 

\begin{equation}
   \vec{u_p} = \left(1 + \frac{(a/2)^2}{6} \nabla^2 \right) \vec{u_f}(\vec{x_p}).
   \label{eq:faxen_vel}
\end{equation}

\begin{equation}
   \vec{\omega_p} = \vec{\omega}(\vec{x_p}).
   \label{eq:faxen_rot}
\end{equation}

where $\vec{u_f}(\vec{x_p})$ is the fluid velocity evaluated at the particle position, and $\vec{u_p}$ is the particle velocity~\cite{guazzelli2011physical}. In the presence of spatially varying velocity gradients, such as in a Poiseuille flow, the velocity field yields a non-zero $\nabla^2 \vec{u_f}$. As a result, a finite-size correction term appears, introducing a dependence on the particle size. However, this correction term does not exist in the torque expression. Therefore, particle size should not influence the particle's rotation rate. This implies that the particle lags behind the fluid velocity by an amount proportional to $a^2\nabla^2 \vec{u_f}$. This leads to a scaling of the velocity difference $U_{\text{Fax}} \sim ((a/2)/w)^2 U  = (1/4) \alpha^2 U$, where $U$ is the characteristic fluid velocity, and has the same rotation rate as the surrounding fluid. In unidirectional flows, as in Poiseuille flows, the Fax\'en correction cannot account for cross-streamline migration, as it acts parallel to the flow velocity. However, in the presence of curved streamlines, the observed velocity difference is not colinear with the local flow direction, which can result in cross-streamline migration. In our system, which features complex spatially varying velocity gradients and curved streamlines, the Fax\'en correction is shown in the inset of Figure~\ref{fig:figure7} (c). The corrections are too small to account for the full magnitude of the velocity difference as they represent less than $\approx0.5\%$ of $|\vec{u_f}|$ as shown in the inset of Figure~\ref{fig:figure7} (c).

Figure~\ref{fig:figure8} (c) shows the difference in rotation rate between particles and fluid along the $x$-direction $\mathcal{R} = (\vec{\omega_p}- \vec{\omega})\cdot \vec{e_x}$. For trajectories that do not interact with the vortex, $\mathcal{R}$ is close to zero or slightly positive, indicating that size has little to no effect. However, for trajectories that spiral around the vortex, $\mathcal{R}$ becomes negative. In the single-phase flow, vorticity is at its maximum around the core. However, the negative $\mathcal{R}$ in these regions indicates that the particle's rotation rate is slower than the fluid's.

Outside the vortex, fluid vorticity decreases and becomes negative (Figure~\ref{fig:figure2}). In this region, particles rotate slightly faster than the surrounding fluid. As stated earlier,  the Fax\'en law does not predict any difference between particle and fluid rotation rates (see equation~\ref{eq:faxen_rot}). In conclusion, if particle dynamics were governed solely by size effects, particle rotation should not be influenced. 

To summarize, we have shown that the Fax\'en correction is too small to account for the observed velocity difference between particles and the fluid. Moreover, the rotation rate difference does not vanish. Therefore, the finite-size and flow curvature effects arising from the Fax\'en law at vanishing Reynolds number do not entirely capture the dynamics of particles in our vortex. This strongly implies that an additional mechanism is required to explain the particle dynamics. In the following, we discuss the role of inertial effects.
\bigbreak
It is known that fluid inertia can generate a lift force acting on particles suspended in a flow. This force acts in the direction of fluid velocity gradients and therefore allows particle drift across streamlines. When the particle Reynolds number $Re_p$ is small, as in the present study, this lift force can be obtained using singular perturbation methods based on the solutions for vanishing Reynolds number~\cite{stone2000philip}. In the absence of walls, the lift force and the resulting inertial migration velocity can then be estimated using scaling arguments. For neutrally buoyant particles, the leading order force singularity is a stresslet, with magnitude of $S\sim \eta a^3 G$, where $G \sim U/W$ is the global shear rate.  In the absence of fluid inertia (i.e., Re$\ll 1$), this stresslet does not affect the motion of the particle in an unbounded domain. However, when fluid inertia is accounted for, this stresslet induces an inertial flow field that becomes comparable in magnitude to viscous effects at a characteristic distance of $a$\text{Re}$_p^{-1/2}$ from the particle. This stresslet-induced inertial flow field disturbance transports the particle across the streamlines. The resulting inertial migration velocity can be estimated by the balance between inertial forces and stresslet-induced volume forces, which scales as $U_L\sim a G$\text{Re}$_p$ \cite{schonberg1989inertial,hogg1994inertial,asmolov1999,stone2000philip}.

In the presence of walls, the effect of confinement on the inertial migration of neutrally buoyant particles has been addressed in the literature for simple geometries. In the limit of small confinement, defined as $\alpha = a/w \ll 1$
the matched asymptotic expansions of~\cite{schonberg1989inertial} and~\cite{asmolov1999} for a 2D Poiseuille flow with mean velocity $U$ show a scaling similar to the unbounded case. 

In this regime, the corresponding inertial cross-streamline migration velocity scales as: 
\begin{equation}
    U_L\sim  a G\text{Re}_p \sim U \alpha \text{Re}_p \sim  U \alpha^3\text{Re}.
    \label{eq:scaling_UL}
\end{equation}

A more recent numerical and theoretical work by Hood \textit{et al.} \cite{hood2015inertial} shows that this scaling is valid in the range $\alpha \lesssim 0.1$.

In the present setup, the confinement is small, with $\alpha = a/w = 0.018-0.127$, as well as the particle Reynolds number, Re$_p \lesssim 1$. In addition, the spherical particles spiraling in the cross-slot region remain far from the channel walls. We can therefore expect the particles to evolve within the asymptotic regime studied by~\cite{schonberg1989inertial,asmolov1999}, where the inertial migration velocity scales as given by Equation 
\eqref{eq:scaling_UL}.

The relative velocity arising from Fax\'en law \eqref{eq:faxen_vel} scales as discussed above as $U_{\text{Fax}} = (1/4) \alpha^2 U$.
If the inertial cross-streamline migration velocity $U_L$ scales as $\alpha^3$\text{Re}$U$ 
then the ratio of the two velocities should scale as $U_L/U_{\text{Fax}}\sim 4 \alpha$Re. 
Using the mean velocity difference $\mathcal{U}$ and the mean Fax\'en velocity  measured in our simulations with 80 $\mu$m and 140 $\mu$m spheres (see Figure \ref{fig:figure7} (c)), we obtain:

\begin{equation}
    \frac{\mathcal{U}_{80}}{U_{\text{Fax},80}} = \frac{0.02}{0.0011} = 18.2, 
\end{equation}
and
\begin{equation}
    \frac{\mathcal{U}_{140}}{U_{\text{Fax},140}} = \frac{0.04}{0.0015} = 26.7.
\end{equation}

Indeed these ratios are close to $4 \alpha_{80}\text{Re} = 4 \times 0.073\times 56 = 16.5$ and $4 \alpha_{140}\text{Re} = 4 \times 0.127\times 56 = 28.5$. The close agreement between the measured and expected velocity ratios confirm the $a^3 Re$ scaling of the inertial migration velocity. This is in qualitative agreement with the observations described in section~\ref{sec:statistics} that particle distributions depend strongly on particle size but only weekly on the Reynolds number in the range of parameters tested.

However, even though the particles remain far from the walls, where the $a^3$ scaling of the migration velocity should not depend much on the domain geometry, the steady flow field in the cross-slot region is significantly more complex than the symmetric square-duct and Poiseuille flows for which the theoretical framework is well established. To our knowledge, no theoretical studies currently provide a scaling law of the migration velocity as a function of particle size in vortical flows. 
Given that, near the vortex core, the streamlines closely resemble an axisymmetric Burgers vortex flow, a theoretical analysis on inertial migration velocity in such a canonical flow could be useful to confirm the $a^3$ scaling we seem to observe in our study. In addition, investigating the relative rotation between particles and the surrounding flow could be particularly interesting in these vortical flows. Such an analysis would be relevant to a variety of situations where particles migrate across vortices, including turbulent flows.

\section{Conclusion} \label{sec:conclusion}
In this paper, we performed experiments and numerical simulations to study particle transport in a single well-controlled vortex flow. In the low confinement regime explored here, particles follow two types of trajectories, depending on their initial position: they either enter the vortex or completely bypass it. From the perspective of the Core Vortex Plane (CVP), particles interact with the vortex only if they enter between the separatrix streamlines. From the Stretched Vortex Plane (SVP) perspective, particles that enter near the CVP revolve longer around the stagnation point before exiting this plane and proceeding towards the outlets. In contrast, particles entering farther from the CVP perform larger spirals around the rotation axis or are directly advected into one of the outlets. A striking result we obtained is that these particles tend to follow a self-similar decaying motion in the cross-slot. This behavior follows the same power-law scaling as the streamlines in a Burgers vortex. This finding demonstrates that their motion is primarily governed by the balance between viscous diffusion and vortex stretching.

However, as particle size is increased, particles tend to deviate from this self-similar trend. The attenuation of their motion becomes slower than that predicted for a Burgers vortex, indicating that larger particles are progressively pushed away from the vortex core. Experimental statistics of particle presence in the CVP confirm this trend. Larger particles do not reach the vortex core at all. 

Through simulations, we were able to show that this deviation from flow streamlines is driven by fluid inertia. Fluid inertia gives rise to a lift force acting on the particles, which enables drifting across streamlines, resulting in a migration velocity that grows as particle size is increased. From our data, we are able to show that, in the case of our vortical flow field, the migration velocity scales with the particle diameter $a$ as $a^3$, or more precisely as $(a/w)^3 Re$. 
It is important to note that, in agreement with the small particle Reynolds numbers obtained in our weakly confined system and the intermediate flow Reynolds numbers $\text{Re}$ considered here, particle migration velocities represent only a few percent of the mean flow velocity. However, the cumulative effect of this gradual drift along particle trajectories leads to noticeable effects. In particular, it results in spatially varying particle concentrations and a noticeable depletion of particles in the vortex core.

Surprisingly, in the complex and spatially varying vortical flow field investigated in this study, we recover a scaling for the migration velocity similar to that predicted for unidirectional Poiseuille flows. However, no theoretical studies are currently available to predict the exact local migration velocities in such complex vortical flow. As vortices are ubiquitous in nature, it would be helpful to conduct theoretical work to better understand the observed particle migration velocity. 
The vortical flow examined in this study closely resembles a Burgers vortex. These stretched vortices can be considered fundamental building blocks of more complex flow fields, including aspects of turbulence. Obtaining scaling laws for migration velocity and lift force for a canonical flow like a Burgers vortex might pave the way for improved understanding of particle motion in turbulent flows.

We briefly discussed in the introduction how non-Newtonian fluids affect the transport of particles in a straight channel. Extending this investigation to viscoelastic fluids, coupled with flow inertia, within our geometry would be of particular interest. It has been shown that introducing polymers will decrease the intensity of the vortex. However, it also introduces additional stresses, such as the normal stress difference, which depends on flow gradients, fluid viscosity, and polymer properties~\cite{burshtein2017inertioelastic,clarke2025movement}. 
This additional polymeric stress is expected to be particularly intense in the vortex core. As a result, particles may be forced to migrate toward regions of lower normal stress, away from the rotation axis. Performing experiments that introduce elasticity into the flow would provide new insights and could potentially reveal a new way to control the migration of particles around a vortex.

Finally, following this study with spherical particles, one can wonder how the particle shapes affect the particle transport in such flows. Fibers and deformable particles are prevalent in natural environments and industrial processes. Introducing an additional degree of freedom, such as orientation for anisotropic particles or shape changes for deformable ones, is expected to give rise to new migration behaviors. 
The behavior of non-spherical particles has been previously studied in simple flow configurations, such as a Poiseuille flow. These studies show that fibers tend to align with flow gradients, particularly with the vorticity. Extending such investigations to more complex flows through experiments and numerical simulations would allow the study of preferential orientation and its impact on 
migration. This would provide a natural extension of the results presented in this paper.

\section*{ACKNOWLEDGMENTS}
We thank Timm Krüger, Clément Bielinski, Mahdi Davoodi and Howard Stone for fruitful discussions. We thank the Institut Pierre-Gilles de Gennes' platform for their help and the equipment used to fabricate the PDMS device.

AL and NB acknowledge support from the ERC Consolidator Grant
PaDyFlow under Grant agreement 682367. We thank Institut Pierre-Gilles de Gennes (Investissements d’avenir ANR-10-EQPX-34) and MA acknowledges IdEx (ANR-18-IDEX-0001).

The computations were enabled by resources provided by the National Academic Infrastructure for Supercomputing in Sweden (NAISS), partially funded by the Swedish Research Council through grant agreement no. 2022-06725.

SJH and AQS gratefully acknowledge the support of the Okinawa Institute of Science and Technology Graduate University (OIST) with subsidy from the Cabinet Office, Government of Japan, and also funding from the Japan Society for the Promotion of Science (JSPS, Grant No. 24K07332 and 24K00810).

\bibliography{bibliography}
\bibliographystyle{unsrt}

\end{document}